# Attention-based 3D CNN with Multi-layer Features for Alzheimer's Disease Diagnosis using Brain Images*

Yanteng Zhang, Qizhi Teng, *Member, IEEE,* Xiaohai He, *Member, IEEE,* Tong Niu, Lipei Zhang,
Yan Liu, and Chao Ren, *Member, IEEE*

*Abstract*—Structural MRI and PET imaging play an important role in the diagnosis of Alzheimer's disease (AD), showing the morphological changes and glucose metabolism changes in the brain respectively. The manifestations in the brain image of some cognitive impairment patients are relatively inconspicuous, for example, it still has difficulties in achieving accurate diagnosis through sMRI in clinical practice. With the emergence of deep learning, convolutional neural network (CNN) has become a valuable method in AD-aided diagnosis, but some CNN methods cannot effectively learn the features of brain image, making the diagnosis of AD still presents some challenges. In this work, we propose an end-to-end 3D CNN framework for AD diagnosis based on ResNet, which integrates multi-layer features obtained under the effect of the attention mechanism to better capture subtle differences in brain images. The attention maps showed our model can focus on key brain regions related to the disease diagnosis. Our method was verified in ablation experiments with two modality images on 792 subjects from the ADNI database, where AD diagnostic accuracies of 89.71% and 91.18% were achieved based on sMRI and PET respectively, and also outperformed some state-of-the-art methods.

## I. INTRODUCTION

The increasing aging of the global population has led to an increase in the number of patients with Alzheimer's disease (AD). As the most common neurodegenerative disease, AD is causing a progressive damage to the memory and cognitive functions of patients [1]. Mild cognitive impairment (MCI) is a preclinical phase of the AD, which is defined as cognitive decline but that does not interfere notably with activities of daily life, most of MCI will progressively transform into AD. Therefore, the early prediction and screening of AD is of great importance to patients. Nowadays, AD diagnosis is still a challenge in clinical practice, it depends heavily on neurological assessment, behavioral testing and neuroimaging.

Neuroimaging provides a more powerful basis for screening and diagnosis of AD, among which structural magnetic resonance imaging (sMRI) and 18 fluorodeoxyglucose positron emission tomography (FDG-PET) are the most common brain image technologies for the AD auxiliary diagnosis [2]. sMRI can provide brain structural information, such as gray matter (GM), cerebrospinal fluid and other morphological information. Meanwhile, PET assesses the glucose uptake of the brain to indicate tissue metabolic activity by injection of tracers. Recently, many studies [3][4] proposed computer-aided diagnosis methods for AD based on brain images, such as based on sMRI or PET.

Due to the development of artificial intelligence and the increase of available medical data, deep learning has become a major research method of AD computer-aided diagnosis recently. CNNs, as a deep learning method with strong nonlinear data representation capability, have achieved commendable achievements in exploring AD diagnosis. Most CNNs are designed for 2D image recognition, so some studies trained network by extracting 2D slices, but this loses the spatial information features of brain image. Many 2D slices-based methods resulted in testing with inaccurate or excessive accuracy due to data leakage [5][6]. In contrast, 3D CNN enables effective brain image feature extraction with higher efficiency and accuracy. Korolev [7] proposed a depth 3D CNN method by learning discriminant features based on sMRI to predict AD. However, some 3D CNN baseline models lacking the combination with clinical knowledge and were not customized to model the feature representations of brain image. During the progression of health to AD, the structural and metabolic characteristics of the brain show different changes. The structural changes mainly include atrophy of cerebral cortex, but not limited to hippocampus and amygdala regions. This changes in multiple regions also exist in metabolic characteristics, including the glucose metabolic reductions in the posterior cingulate gyri and the parietal lobule [8][9]. This is one reason why clinical diagnosis through visual analysis of brain images still has limitations.

In CNN-based methods, combining attention mechanism can achieve improved performance for disease diagnosis [10], but for brain images, the effect of attention mechanism is still limited, some CNN models cannot effectively extract features. In order to diagnose AD more accurately, we designed an end-to-end framework of 3D CNN based on brain images. Specifically, for the knowledge of brain images, our method

*Research was partly supported by the National Natural Science Foundation of China under Grant 62271336, Grant 62211530110, the Chengdu Major Technology Application Demonstration Project (Grant 2019-YF09-00120-SN), the Key Research and Development Program of Sichuan Province (Grant 2022YFS0098).

Yanteng Zhang, Qizhi Teng, Xiaohai He, Tong Niu and Chao Ren are with the College of Electronics and Information Engineering, Sichuan University, Chengdu 610065, China (e-mail: yntn32@outlook.com, qzteng@scu.edu.cn, hxh@scu.edu.cn, tongtong@stu.scu.edu.cn, chaoren@scu.edu.cn. Corresponding author e-mail: hxh@scu.edu.cn).

Lipei Zhang is with the Department of Applied Mathematics and Theoretical Physics, University of Cambridge, Cambridge CB3 0WA, UK (email: lz452@cam.ac.uk)

Yan Liu is with the Department of Neurology, the Third People's Hospital of Chengdu, Chengdu 610031, China (e-mail: 408521577@qq.com).

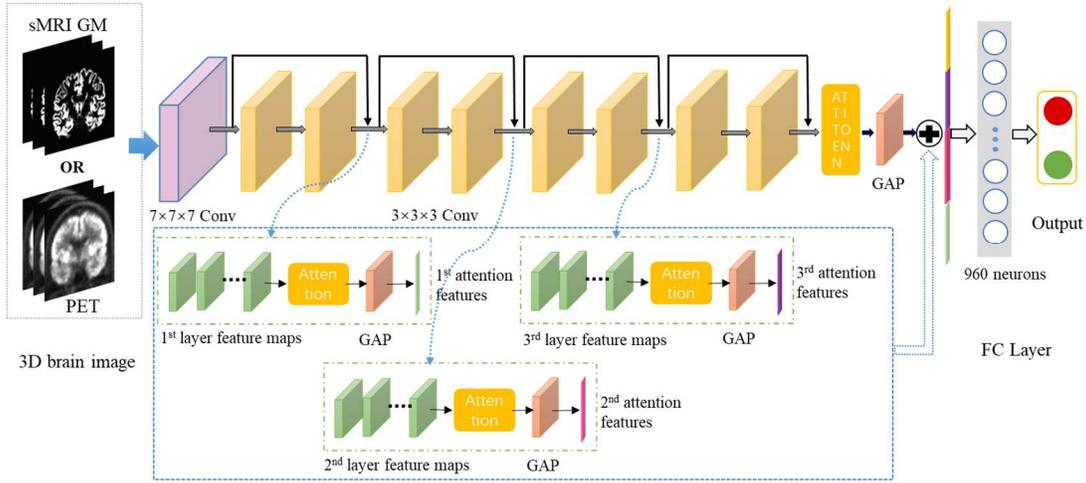

Figure 1. The framework of our network is an attention-based 3D CNN with multi-layer features.

integrates low-level features and high-level features in the last fully connected layer under the effect of attention module, so that the subtle changes in brain image can be further captured to achieve better diagnostic performance. In this work, our main contributions can be summarized as follows: (1) We proposed an end-to-end 3D CNN framework that achieves AD diagnosis without a complex feature extraction and selection process. (2) Our network can focus on important brain regions relevant to disease diagnosis, and we provided interpretability of model via attention maps. (3) Our method is validated on two different modalities with sMRI and PET images, achieved remarkable diagnostic performance for both AD diagnosis and MCI conversion prediction, and outperformed some other deep learning-based methods.

## II. DATA AND PREPROCESSING

In this work, all data we used are obtained from the Alzheimer's Disease Neuroimaging Initiative (ADNI) baseline dataset (adni.loni.usc.edu). We obtained T1 weighted sMRI and FDG-PET images of AD, MCI and NC (Normal Control). We further divided MCI subjects into pMCI and sMCI. The MCI subjects developed AD within 3 years is pMCI, and the MCI subjects did not convert to AD is sMCI. The above image data included 215 AD, 246 NC, 211 sMCI and 120 pMCI. All sMRI images are preprocessed with AC-PC correction and affine-registered to MNI152 space by SPM tool, then extract GM tissue from the preprocessed sMRI image. For PET image, they are affine registered according to the subject's corresponding sMRI. The image resolution after preprocessing is 121×145×121. All images in each modality are not from the same subject to avoid data leakage [5].

## III. METHODS

Currently, many research use the CNN baseline model to directly act on brain images for feature extraction to achieve AD diagnosis. However, the differences between brain images of cognitive impairment are in many cases subtle. In this work, we designed an end-to-end 3D CNN framework based on ResNet [11] architecture for AD diagnosis, which integrates the features from multi-layer under the effect of attention mechanism to better capture differences among brain images. The proposed network framework is shown in Figure 1.

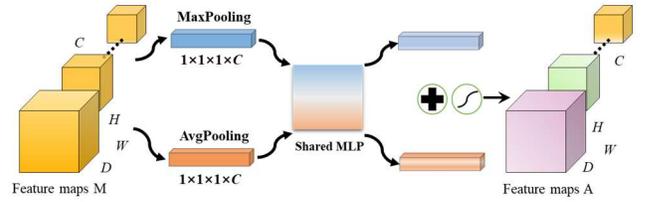

Figure 2. The architecture of our attention module.

### A. 3D CNN

Different from traditional methods of manually extracting features such as cerebral cortex thickness and brain volume, 3D CNN can directly act on brain images to learn general features. In order to effectively encode the spatial information in 3D brain images, we utilize 3D ResNet18 as the architecture, which is mainly composed of four convolutional layers, pooling layer, fully connected layer and softmax layer. In the convolutional layers, the size of kernel is 3×3×3. The number of filters in convolutional layers is 64, 128, 256, 512.

### B. Attention Module

In the neuroimaging manifestations of AD, there are usually morphological or metabolic changes in multiple brain regions. In order to pay more attention to brain regions in features extraction, we designed the attention mechanism [12] to adapt to 3D brain images. The feature maps generated from the convolutional layer are fed to the attention module. The input feature maps are expressed as $M = [M_1, \cdots, M_C]$, where $M_i \in R^{H \times W \times D}$ ($i$=1, 2, ..., C) represents the feature map of the $i$-th channel, and C represents the number of channels. Then, we perform cross-channel average pooling and max pooling on M to generate two feature channels, $M_{avg}$ and $M_{max}$ respectively. The $M_{avg}$ and $M_{max}$ share the same multi-layer perceptron to learn the dependency between channels. Finally, the weight coefficients are obtained by the sigmoid function as the nonlinear activation, and calculate the final attention maps A. In this way, some key feature maps of brain image in A are attentioned under the effect of weights. The architecture of attention module is shown in Figure 2.

## C. Multi-layer Features Concatenation

Different from other diseases, cognitive disorders do not have specific diseased regions in imageology. Moreover, the spatial information of brain image is complex, and the differences between individuals are not obvious. To consider these characteristics, we integrate low-level features and high-level features of convolutional layer in the last fully connected layer in order to reduce information loss in feature extraction.

Throughout the network, in order not to affect the acquisition of raw low-level features and high-level features of brain image, attention module is not used directly in the convolutional layers, but acts independently on the feature maps of each convolutional layer. The raw feature maps in the network continue into the next convolutional layer. Global average pooling (GAP) is performed for the attention-based features obtained after each layer, then the attention-based multilayer features obtained after GAP are concatenated. At the end, a fully connected layer followed by a softmax layer. All the features are flattened into a one-dimensional vector of 960, and the output dimension is the number of categories.

## IV. EXPERIMENTS AND RESULTS

### A. Experimental Settings

We performed two types of classification tasks, in addition to AD detection, achieving the MCI conversion prediction is important for the early treatment of AD. The division ratio of training, validation and test set is 70%:15%:15%. We performed experiments on two sub-datasets for final result by crossing the test and validation sets and evaluated the performance by accuracy, sensitivity, specificity, and AUC.

Our experiments are implemented on PyTorch platform with a GPU NVIDIA P100 16G. We use the Adam algorithm as the optimizer, the cross-entropy loss function is used to optimize the network. The parameters in network training used are as follows, batch size is set to 6, learning rate is initially set to 1e-4, and after 30 epochs every 10 epochs are halved until 5e-6, and the model trained for 70 epochs. The convolution kernel is initialized randomly in AD prediction and then we use the network parameters learned to initialize the training network for MCI conversion prediction.

### B. Results and Discussions

In this section, we performed ablation experiments and analyses based on sMRI GM and PET images respectively, and discuss the impact of attention mechanism on the diagnosis of AD. The results of the sMRI and PET based classification tasks are summarized in Table I.

First, 3D ResNet18 is used as the baseline model for the experiments. When only 3D CNN baseline model was used, the accuracy based on sMRI in AD vs. NC classification reached 86.03%, while the accuracy based on PET reached 87.50%. Under the same baseline, the performance based on PET is better than that of sMRI, also in the sMCI vs. pMCI. Secondly, we combined attention module after the first convolutional layer on ResNet18 architecture (AResNet18). From the classification results in Table I, we can see that the diagnostic performance of the model is slightly improved by combining the attention mechanism but not much. We think this is related to the complex characteristics of brain image. Although some studies have proved that the hippocampus, amygdala in sMRI and cingulate gyri, parietal lobule in PET have changed, it's not very obvious changes to the whole brain. Finally, the results based on our proposed method were improved further. In the AD vs. NC classification, the accuracy based on sMRI reached 89.71%, while the PET-based reached 91.18%. For the prediction of sMCI vs. pMCI, the accuracy based on sMRI and PET is 74.56% and 75.44%. In addition, we achieve an accuracy of 83.71% and 84.83% for AD vs. MCI diagnosis based on sMRI and PET, respectively. In above results, the accuracy of our method based on PET is better. This result is consistent with the established clinical knowledge. PET can detect the functional brain changes and specific pathologies of AD at the early stage than sMRI.

3D Grad CAM can provide better interpretability of AD diagnostic models. We applied 3D Grad CAM to our models to show some important regions of interest to network. Figure 3 shows the heat maps of two convolutional features under MRI GM. As can be seen from the feature maps of both convolutional layers, our attention is mainly focused on the cerebral cortex and hippocampal regions. Figure 4 shows the heat maps of attention to PET, where posterior cingulate gyri, parietal lobule and precuneus are marked as some important regions. The regions are consistent with medical clinical variations [9][13]. While the low-level and high-level features in our network reflect detailed and global feature effects, which we combine to achieve better diagnostic performance.

Furthermore, we compared with several state-of-the-art methods based on ADNI dataset in Table II. Our proposed method shows a better diagnostic performance. Currently, the diagnostic effect of MCI conversion prediction is not ideal, due to there are subtle morphological differences between sMCI and pMCI, which is also a difficult issue in clinical diagnosis. With the accumulation of medical data, multimodal approaches offer more possibilities for accurate diagnosis of AD. The improvement of multimodal performance in some approaches is limited or even the effect is opposite [4]. Further exploration is needed to know how to combine clinical knowledge and effectively use multimodal data for AD diagnosis, that is the trend and focus in future work.

TABLE I. THE CLASSIFICATION RESULTS OF ABLATION EXPERIMENTS BASED ON sMRI AND PET, RESPECTIVELY

| Methods | AD vs. NC | | | | sMCI vs. pMCI | | | |
|---|---|---|---|---|---|---|---|---|
| sMRI | ACC | SEN | SPE | AUC | ACC | SEN | SPE | AUC |
| ResNet18 | 86.03 | 85.94 | 86.11 | 86.64 | 70.18 | 63.64 | 74.86 | 68.96 |
| AResNet18 | 87.50 | 86.93 | 88.89 | 87.42 | 72.81 | 70.46 | 74.29 | 72.37 |
| OURS | 89.71 | 89.71 | 90.28 | 89.67 | 74.56 | 65.91 | 80.0 | 72.96 |
| PET | ACC | SEN | SPE | AUC | ACC | SEN | SPE | AUC |
| ResNet18 | 87.50 | 89.06 | 86.11 | 87.59 | 71.93 | 63.64 | 77.17 | 70.39 |
| AResNet18 | 88.98 | 90.63 | 87.50 | 88.57 | 73.68 | 59.09 | 82.86 | 70.97 |
| OURS | 91.18 | 90.63 | 91.67 | 91.15 | 75.44 | 63.64 | 85.71 | 72.40 |

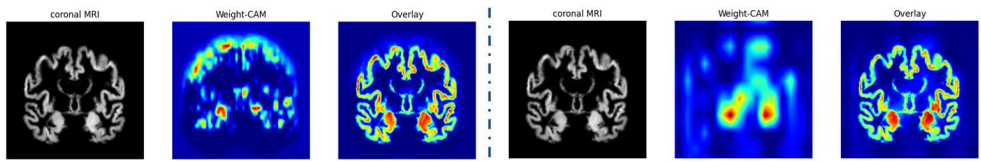

Figure 3. Coronal views of brain MRI GM image, Grad CAM weights, and visual interpretation of the heat maps. We enumerate the application of Grad Cam to the first convolutional layer (left side) and the third convolutional layer (right side) to show the visualization of attention maps.

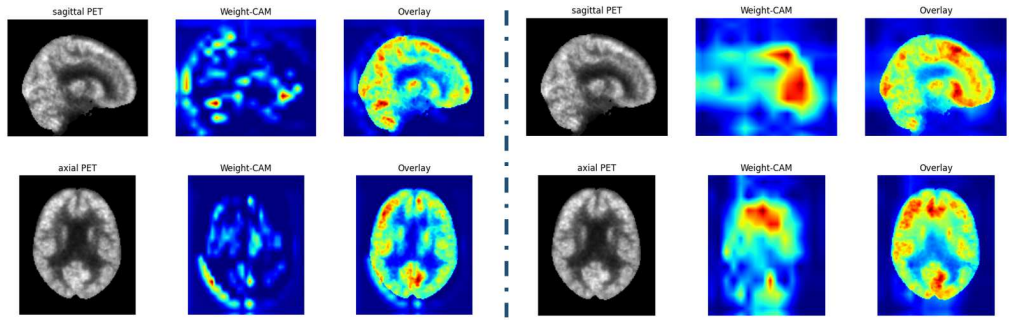

Figure 4. Sagittal and axial views of brain PET image, Grad CAM weights, and visual interpretation of the heat maps. We enumerate the application of Grad Cam to the second convolutional layer (left side) and the fourth convolutional layer (right side) to show the visualization of attention maps.

TABLE II. THE CLASSIFICATION RESULTS IN DEEP LEARNING-BASED STUDIES BASED ON ADNI DATASET

| Study | Method | Image | AD vs. NC | | sMCI vs. pMCI | |
|---|---|---|---|---|---|---|
| | | | ACC | AUC | ACC | AUC |
| Lin [14] | 3DCNN | MRI | 0.820 | 0.818 | 0.728 | 0.683 |
| Marla[4] | 3DCNN | MRI | 0.866 | - | - | - |
| Wen [6] | 3DCNN | MRI | 0.890 | - | 0.740 | - |
| Xiao[15] | GCN | MRI | 0.869 | - | - | - |
| OURS | 3DCNN | MRI | **0.897** | **0.898** | **0.745** | **0.729** |
| Xing[3] | 3DCNN | PET | 0.866 | 0.886 | - | - |
| Lin [14] | 3DCNN | PET | 0.888 | 0.871 | 0.728 | 0.683 |
| Marla[4] | 3DCNN | PET | 0.905 | - | - | - |
| OURS | 3DCNN | PET | **0.912** | **0.910** | **0.754** | **0.724** |

## V. CONCLUSION

In this work, we strictly evaluated the diagnostic performance of AD based on sMRI and PET images through ablation experiments, the PET-based was more advantageous than sMRI. PET can better capture the specific pattern of AD neurodegeneration than sMRI in the early progress. In general, the results showed that our method can diagnose AD more effectively, cause the subtle differences among images can be further learned through the attention-based multi-layer features fusion. Our method may provide a reference for diagnosis based on other medical images. However, there are still many challenges for MCI conversion prediction task, which is also the focus of the research in the future.

## ACKNOWLEDGMENT

Data used in preparation of this article were obtained from the Alzheimer's Disease Neuroimaging Initiative (ADNI) database. As such, the investigators within the ADNI contributed to the design and implementation of ADNI and/or provided data but did not participate in analysis or writing of this report. More details can be found at adni.loni.usc.edu.